\documentclass[fleqn,usenatbib]{mnras}

\usepackage{newtxtext,newtxmath}

\usepackage[T1]{fontenc}

\DeclareRobustCommand{\VAN}[3]{#2}
\let\VANthebibliography\thebibliography
\def\thebibliography{\DeclareRobustCommand{\VAN}[3]{##3}\VANthebibliography}

\usepackage{graphicx}	
\usepackage{amsmath}	
\usepackage{xcolor}

\title[IA contamination to growth rate]{Impact of intrinsic alignments on clustering constraints of the growth rate}

\author[Zwetsloot \& Chisari]{
Karel Zwetsloot$^{1}$ and Nora Elisa Chisari$^{1}$\thanks{E-mail: n.e.chisari@uu.nl}.
\\
$^{1}$Institute for Theoretical Physics, Utrecht University, Princetonplein 5, 3584 CC, Utrecht, The Netherlands.
}

\date{Accepted XXX. Received YYY; in original form ZZZ}

\pubyear{2022}

\begin{document}
\label{firstpage}
\pagerange{\pageref{firstpage}--\pageref{lastpage}}
\maketitle

\begin{abstract}
Intrinsic alignments between galaxies and the large-scale structure contaminate galaxy clustering analyses and impact constraints on galaxy bias and the growth rate of structure in the Universe. This is the result of alignments inducing a selection effect on spectroscopic samples which is correlated with the large-scale structure. In this work, we quantify the biases on galaxy bias and the growth rate when alignments are neglected. We also examine different options for the mitigation of alignments by considering external priors on the effect and different probe combinations. We find that conservative analyses that restrict to $k_{\rm max}=0.1$ Mpc$^{-1}$ are not significantly affected. However, analyses that aim to go to higher wave numbers could evidence a significant contamination from alignments. In those cases, including a prior on alignment amplitude, or combining clustering with the position-intrinsic shape correlation of galaxies, can recover the same expected constraining power, or even inform bias and growth rate measurements. 
\end{abstract}




\section{Introduction}

Surveys of the extragalactic Universe aim to constrain the cosmological model by measuring the growth rate of structure at different epochs. This is quantified by the parameter 
\begin{equation}
    f=\frac{d\ln D}{d\ln a}
\end{equation}
where $D$ is the linear growth rate of the matter density contrast, and $a$ is the scale factor. Deviations from general relativity can be identified from the scaling of this parameter with $a$ or with redshift, $z$. 

Constraints on $f$ are obtained thanks to the sensitivity of the peculiar velocities of galaxies to the growth of structure. These peculiar velocities have an impact on the observed galaxy redshift. This translates into anisotropic ``redshift-space distortions'' (RSDs, \citealt{Kaiser87}) on the clustering statistics (see \citealt{Percival09} for a review), which can be measured with great precision \citep[e.g.][]{Alam21}. In practice, such clustering statistics constrain the combination of $f\sigma_8$ (growth rate and amplitude of matter fluctuations). This can also be extrapolated from the best-fit $\Lambda$ CDM model of the Cosmic Microwave Background (CMB) data from the {\it Planck} mission \citep{Planck}. The current literature displays some signs of tension with respect to the growth predictions extrapolated from the CMB. Many authors report significantly discrepant values ($2-3\sigma$) for $f\sigma_8$ \citep[e.g.][]{damico20,Ivanov21,Chapman21,Kobayashi22,Yuan22,Zhai22}, while others are in better agreement with {\it Planck} \citep[e.g.][]{Beutler12,Reid14,Alam17,deMattia21,Bautista2021,Lange22,Chen22,Philcox22,Zhang22}. The analyses are performed under a variety of conditions: different sample selections and redshifts, different scales and thus modelling methods. If a low $f\sigma_8$ is confirmed, it could imply deviations from General Relativity \citep{Jain08}.

\citet{Hirata09} demonstrated that certain measurements of the growth rate are expected to be contaminated by intrinsic alignments. Galaxy intrinsic alignments are physical correlations between galaxy shapes and orientations and the large-scale structure, caused, for example, by tidal interactions \citep{Catelan01}. The contamination of RSD measurements comes about because the orientation of galaxies often determines their detection probability. Galaxies aligned along the line of sight are more likely to be detected than those aligned across the line of sight. (Detections based on model or Petrosian magnitudes are not affected; but cases where magnitudes come from isophotal or fixed-aperture measurements will suffer from contamination because a larger proportion of their flux lies within a fixed aperture.)

\citet{Martens18} attempted to measure the impact of alignments on RSD directly by comparing clustering measurements in two samples of galaxies selected by their sizes. According to \citet{Hirata09}, the observed radius of a galaxy depends on its orientation. Because this orientation is correlated with the tidal field, the observed size is biased with respect to the original one. The sizes of elliptical galaxies are measured with respect to a scaling relation known as the ``Fundamental Plane'' \citep{Djorgovski87}. Using data from the Baryon Oscillation Spectroscopic Survey \citep[ BOSS,][]{Dawson13}, \citet{Martens18} assumed that populations above or below the plane would have different orientations with respect to the line of sight. They found a $2-3\sigma$ difference in the clustering properties of the two samples, which they ascribed to intrinsic alignment bias. A key assumption behind the work of \citet{Martens18} is that size deviations from the Fundamental Plane can be solely attributed to the impact of alignments on the observed radii of galaxies. The detection is also contingent on combining the two main spectroscopic samples of the BOSS survey and not in each sample alone. \citet{Singh21} pointed out that the deviations from the Fundamental Plane can also receive significant contributions from other physical processes or observational systematics, complicating the picture and suggesting that intrinsic alignments are not inducing any detectable selection effects on the BOSS samples analysed by \citet{Martens18}. This was confirmed in \citet{Zhang22}, where the contamination of intrinsic alignments was explicitly modelled when obtaining cosmological constraints from galaxy clustering (see their Appendix D). At what level one should expect the effect proposed by \citet{Hirata09} to be present in upcoming survey data remains unclear, and the effect is probably dependent on the galaxy sample. \citet{Agarwal21} demonstrated that the deterioration of constraints on the growth rate can be severe for upcoming surveys such as {\it Euclid} if this type of systematics are neglected.

In this work, we explore different scenarios for the impact of intrinsic alignments on the growth rate. We consider a case where the effect of intrinsic alignments is completely neglected, and other cases where intrinsic alignment observables are used in combination with galaxy clustering measurements. We focus on luminous red galaxies, for which there is ample observational evidence supporting a linear alignment with the tidal field on large scales \citep{Mandelbaum06,Hirata07,Joachimi,Singh,Johnston,Fortuna21}. We perform our analysis in a set-up that matches the expected survey properties of the 4MOST spectroscopic instrument's Cosmology Redshift Survey \citep{4MOST}, assuming it to be complemented by galaxy shapes from the Legacy Survey of Space and Time (LSST) over the same area \citep{LSST}. Nevertheless, our method is general and can be applied to any other survey combination. All in all, we demonstrate that the inclusion of alignment observables can mitigate several $\sigma$ biases on the growth rate, or even improve its constraints.

This paper is organised as follows. Section \ref{sec:formalism} presents the modelling and forecasting formalism and describes the fiducial modelling assumptions. Section \ref{sec:results} summarizes our results. The conclusions are presented in Section \ref{sec:conclusions}. Appendix \ref{app:multi}, provides auxiliary calculations to Section \ref{sec:formalism}.

\section{Formalism}
\label{sec:formalism}

We assume that the cosmological information is encoded in the power spectra between two observable fields, $X$ and $Y$, given by $P_{XY}(k,\mu)$. This is a function not only of the wavenumber but also $\mu=k_z/k$, where $\hat{\bf z}$ is the line of sight direction. We summarise this information in a set of multipoles by expanding the power spectra as
\begin{equation}
    P_{XY}(k,\mu)=\sum_{l=0}^{\infty}P_{XY}^l(k)\mathcal{L}_l(\mu),
\end{equation}
where $\mathcal{L}_l(\mu)$ are Legendre polynomials. Each multipole is then given by
\begin{equation}
    P_{XY}^l(k)=\frac{2l+1}{2}\int_{-1}^{1}d\mu\,P_{XY}(k,\mu)\mathcal{L}_l(\mu).
    \label{eq:multi}
\end{equation}

In principle, all power spectra and multipoles are redshift-dependent. For simplicity, we will work at a fixed mean redshift specified for each survey in Section \ref{sec:fiducial}.

\subsection{Galaxy clustering}
\label{sec:gg}

Galaxy redshifts are affected by large-scale motions \citep{Kaiser87} and their velocities grow with the build-up of structure. This induces an angular dependence to the galaxy clustering power spectrum. In a linear bias model, this is given by
\begin{equation}
    P_{gg}(k,\mu)=(b_g+f\mu^2)^2P_{\delta}(k)
    \label{eq:pkgg}
\end{equation}
where $b_g=\delta_g/\delta$ is the linear galaxy bias, $P_{\delta}$ is the linear matter power spectrum and $f$ is the logarithmic growth rate. 

In the presence of intrinsic alignments, Eq. \ref{eq:pkgg} is modified to \citep{Martens18}
\begin{equation}
    P_{gg}(k,\mu)=\left[b_g-\frac{A}{3}+(f+A)\mu^2\right]^2P_{\delta}(k)
    \label{eq:pkggA}
\end{equation}
where $A$ is the dimension-less amplitude which is related to the strength of alignment of the specific galaxy population, and to how their orientation determines their selection probability. This expression is only valid at linear level \citep{Hirata09}. From Eq. \ref{eq:pkggA}, we see that If alignments are neglected, galaxy clustering surveys are effectively measuring $\tilde{b}_g= b_g-A/3$ and $\tilde{f}=f+A$. To break the degeneracy between these parameters and obtain an unbiased growth rate measurement, we would need an additional source of information on $A$. This could be an intrinsic alignment observable, as we will describe in the following subsections. 

Notice that because we work at a fixed redshift, we will report the contamination on $f$ directly. However, as \citet{Hirata09} points out, if the matter power spectrum is normalized at high redshift by constraints obtained from the CMB, then RSD measurements truly constrain $f(z)D(z)$.

It can be easily verified that for $P_{gg}(k,\mu)$, the only nonzero multipoles are the monopole ($l = 0$), the quadrupole ($l = 2$), and the hexadecapole ($l = 4$). Most of the signal-to-noise is contained in the lowest order multipoles. We will, as it is often done \citep[e.g.][]{Blake18}, neglect the information in the hexadecapole. The expressions for the multipoles are presented in Appendix \ref{app:multi}.

\subsection{Galaxy shapes}

In photometric surveys, the shapes of galaxies are measured in projection on the sky and modelled as ellipses. To describe a shape, we thus need two numbers: the axis ratio ($q=a/b$) and the orientation of the ellipse ($\phi$). We can then estimate the two-component shape as:
\begin{equation}
    (e_+,e_\times)=\frac{1-q^2}{1+q^2}(\cos(2\phi),\sin(2\phi))
\end{equation}
From the theory point of view, the effects that contribute to the measured shapes are: random noise, gravitational lensing and intrinsic (physical) alignments between galaxies and the matter field. The shapes are also most often modelled in Fourier space, transforming $(e_+,e_\times)$ into $(\gamma_E,\gamma_B)$ in analogy with the $E,B$ decomposition traditional of CMB polarisation literature \citep{Kamionkowski,Zaldarriaga}.

In this work, we will focus mostly on the $E-$modes of the intrinsic alignment ($I$) component of the shape, thus equating the observed $\gamma_E^{\rm obs}=\gamma_E^I+\gamma_{\rm rnd}$, where the second term is the noise.
In principle, gravitational lensing also correlates the observed shapes of galaxies with the large-scale density field (responsible for perturbing photon geodesics) and with one another (i.e. two galaxies being lensed by the same structure). However, we omit any gravitational lensing contributions to our observables as \citet{Taruya20} suggested these are negligible in our context. The reasons behind these simplifications are as follows. In measuring cross-correlations of galaxy positions and shapes from spectroscopic data, one would expect to be able to isolate close pairs of galaxies along the line of sight. Such procedure would down-weight the lensing contribution significantly, which peaks when the distance between the source and the observer is twice of the distance between the observer and the lens. In the case of shape-shape correlation, the availability of accurate redshifts is not enough to remove the lensing contribution, since galaxies at the same redshift are lensed by the same structure. However, \citet{Taruya20} estimated this would be negligible in our setup.

We also do not consider intrinsic alignment $B$-modes. These have been shown to be absent at the linear level both in alignment and lensing statistics. In the position-shape statistics, they cannot arise unless there is a breaking of parity. In shape-shape statistics, $B-$modes can arise at the quasi-linear \citep{Blazek,Vlah} or nonlinear level \citep{Hui}. Because we work at the linear level in this work, we do not account for $B-$modes.

\subsubsection{Intrinsic alignments}

Galaxy alignments are thought to be induced by gravitational tides. At lowest order in gravitational observables, the intrinsic shapes of galaxies are proportional to the tidal field \citep{Catelan01} the `linear alignment model'. As a result, shapes become correlated with the density field. In particular, there is a correlation between $E$-modes of galaxy shapes, $\gamma_E$, and the linearly biased galaxy overdensity \citep{Hirata04,Taruya20}, which in the linear alignment model is given by
\begin{equation}
    P_{gE}(k,\mu)=-\tilde{C_1}(1-\mu^2)\left(b_g+f\mu^2\right)P_\delta(k)\label{eq:PgE}
\end{equation}
where $\tilde{C_1}$ is the linear bias of galaxy shapes with respect to the tidal field. Notice that our definition of $\tilde{C_1}$ is slightly different from the more commonly adopted $C_1$ in the literature \citep{Singh}, and that the two are connected by $\tilde{C}_1=A_I C_1{\rho_{\rm crit}\Omega_{\rm m}/D(z)}$, where $\rho_{\rm crit}$ is the critical density of the Universe today and $A_I$ and $C_1$ are constants.
    
In the linear alignment model, the auto-correlation of $E$-modes is
\begin{equation}
    P_{EE}(k,\mu)=\left[\tilde{C_1}(1-\mu^2)\right]^2P_{\delta}(k)\label{eq:PEE}
\end{equation}
 For simplicity, we will work only with the lowest order multipoles of $P_{gE}(k,\mu)$ and $P_{EE}(k,\mu)$, i.e. their monopoles, given in Appendix \ref{app:multi}. In principle, we would also need to include the selection effect from alignments in Eq. (\ref{eq:PgE}). This can be done, for example, following \citet{Singh21}. However, we checked that such a term is subdominant in our results.

\subsubsection{Relation between $A$ and $\tilde{C}_1$}
\label{sec:relateAC1}

Breaking the degeneracy between $b_g$, $A$ and $f$ making use of alignment measurements of $P_{gE}(k,\mu)$ and $P_{EE}(k,\mu)$ is only possible if we have a relation between $A$ and $\tilde{C}_1$. \citet{Hirata09} suggests one can interpret $A$ as a multiplicative relation between the alignment bias, which is referred to as $B$, and a selection bias, $(\eta\chi)_{\rm eff}$:
\begin{equation}
    A=2(\eta\chi)_{\rm eff}B. \label{eq:abc}
\end{equation}
$(\eta\chi)_{\rm eff}$ effectively represents the change in the number of galaxies with respect to a change in their radial orientation. In particular, $\eta$ is the slope of the cumulative luminosity function of the population of observed galaxies, and $\chi$ is a number that depends on the method used for computing fluxes and on the typical surface brightness profiles of the sample. \citet{Hirata09} showed that the selection effects induced by intrinsic alignments on clustering measurements are only sensitive to the combination of $\eta$ and $\chi$. More specifically, $(\eta\chi)_{\rm eff}$ in Eq. (\ref{eq:abc}) represents the effective response over the whole sample of observed galaxies. Notice, moreover, that the relation between alignment amplitude given in Eq. (\ref{eq:abc}) is fully deterministic. 

\citet{Martens18} connect $B$ to the more typical parameters used in observational works \citep[e.g., $C_1$ or $A_{\rm IA}$][]{Singh}. Simply accounting for our different definition of the alignment amplitude, $\tilde{C}_1$, we find that $B=-1.74\tilde{C}_1$. The connection is established by realising that our parameter $\tilde{C}_1$ is also equal to $-b_\kappa$, a parameter used in \citet{Martens18} and originally defined in \citet{Bernstein09}. This shows that the factor of $1.74$ also represents the shear responsivity, i.e. the response of measured galaxy shapes to a shear induced by weak gravitational lensing.

We can then finally write $A$ in terms of $\tilde{C}_1$ as $A=-3.48(\eta\chi)_{\rm eff}\tilde{C}_1$. Notice a negative $A$ implies that neglecting alignments in RSD measurements will result in an inferred lower growth rate according to Eq. (\ref{eq:pkggA}).

\subsection{Forecasting}

We forecast the uncertainty on the growth rate and galaxy bias using the Fisher matrix formalism. To estimate the uncertainties in the cosmological parameters, we need the covariance of the multipoles, given by \citet{Taruya11} as
\begin{eqnarray}
    &{\rm Cov}[P_{XY}^l(k),P_{WZ}^{l'}(k)]=\frac{(2l+1)(2l'+1)}{4}\times\label{eq:cov}\\
    &\,\,\,\,\,\,\,\,\,\,\,\int_{-1}^{1}d\mu\,\mathcal{L}_l(\mu)\mathcal{L}_{l'}(\mu){\rm Cov}[P_{XY}(k,\mu),P_{WZ}(k,\mu)]\nonumber
\end{eqnarray}
where ${\rm Cov}[P_{XY},P_{WZ}]=P_{XW}P_{YZ}+P_{XZ}P_{YW}$ is the usual power spectrum covariance and where we omitted the $k$ and $\mu$ dependencies for simplicity. Notice any auto-spectra include the shot or shape noise contributions: $n_{g}^{-1}$ and $\sigma_\gamma^2n_g^{-1}$, respectively, where $n_g$ is the number density of galaxies and $\sigma_\gamma$ the dispersion in their ellipticities per component. For example, while $P_{EE}$ is an auto-correlation of galaxy shapes and subject to both cosmic variance and Poissonian shape noise, $P_{gE}$ only suffers from cosmic variance and is in practice a much more precise probe of galaxy alignments (see e.g. \citet{Blazek11} for an observational demonstration).

The Fisher matrix of the cosmological or nuisance parameters of interest, in our case $\theta_i=\{b_g,f,\tilde{C}_1\}$, is given by
\begin{equation}
    F_{ij}=\frac{V_s}{4\pi^2}\int_{k_{\rm min}}^{k_{\rm max}}dk\,k^2\,\sum_{a,b}\sum_{l,l'}\frac{\partial P_a^{l}(k)}{\partial\theta_i}[{\rm Cov}^{-1}]|_{ab,ll'}\frac{\partial P_b^{l'}(k)}{\partial\theta_j}
    \label{eq:fisher}
\end{equation}
where $a$ and $b$ run over the different elements of the data vectors, i.e. the multipoles considered, and $V_s$ is the survey volume. The uncertainties in the parameters $\delta\theta_i$ are given by $F_{ii}^{-1/2}$ (assuming perfect knowledge of all other parameters) or $(F^{-1})_{ii}^{1/2}$ (marginalising over all other parameters). By default, we will report the marginalised uncertainties in Section \ref{sec:results}. We will also consider the impact of an external prior on the alignment amplitude. This can be included in the Fisher matrix by simply adding $\sigma_{\tilde C_1}^{-2}$ to the corresponding diagonal element.

We also quantify the constraining power for different probe combinations and settings by calculating the figure of merit \citep{DETF} in the $(b_g,f)$ parameter space. This is given by ${\rm FoM}=\pi/S$, where $S$ is the area of the ellipse corresponding to $68\%$ confidence level constraints on the parameters. In our case, the two parameters of interest are $(b_g,f)$. The larger the figure of merit, the better the constraint in this combination of parameters.

It is also possible to extend this formalism to compute the parameter bias in the case that the data is analyzed with the wrong model \citep{Taylor,Amara}. By analogy with Eqs. (7) and (8) of \citet{Amara}, in our case, the bias in a model parameter, $\theta_i$, is given by
\begin{equation}
    b[\theta_i]=(F^{-1})_{ij}B_j
    \label{eq:bias }
\end{equation}
where 
\begin{equation}
    B_j=\frac{V_s}{4\pi^2}\int_{k_{\rm min}}^{k_{\rm max}}dk\,k^2\,\sum_{a,b}\sum_{l,l'} P^{l,{\rm sys}}_a(k)[{\rm Cov}^{-1}]|_{ab,ll'}\frac{\partial P_b^{l'}(k)}{\partial\theta_j}
    \label{eq:biasvec}
\end{equation}
where $P^{l,{\rm sys}}_a(k)$ is the contribution from the missing systematic in the model. In our case, this is computed as the difference between the multipoles with and without intrinsic alignments in the model.

\subsection{Fiducial model}
\label{sec:fiducial}

We perform our forecasts for the Luminous Red Galaxy sample expected to from the 4MOST Cosmology Redshift Survey \citep{4MOST}. Our set-up follows similar choices to \citet{vanGemeren}. The mean redshift of the targets is expected to be $\langle z \rangle =0.55$ with an average number density of $n_g=2.9\times 10^{-4}$ Mpc$^{-3}$. We assume that shapes for this sample will be delivered by the Legacy Survey of Space and Time \citep[LSST,][]{LSST} over the overlapping are (7,500 sq. deg.). We assume a dispersion in intrinsic ellipticities of $\sigma_\gamma=0.3$. 

The volume of the survey ($\simeq 10^{10}$ Mpc$^3$) determines the minimum wavenumber: $k_{\rm min}=2\pi/V_s^{1/3}$ Mpc$^{-1}$. We vary the maximum wavenumber among from $k_{\rm max}=0.1$ Mpc$^{-1}$ to $k_{\rm max}=0.3$ Mpc$^{-1}$ in order to assess the sensitivity of our results to this choice. 

We use the {\tt Core Cosmology Library v2.3} \citep{Chisari19} to compute the power spectrum. This relies on the Boltzmann code {\tt CLASS} \citep{CLASS}. We adopt the following choice of parameters for a flat, neutrino-less, $\Lambda$CDM cosmology: $\{\Omega_{\rm cdm}=0.27,\Omega_{\rm b}=0.045,h=0.67,A_s=2.1\times 10^{-9},n_s=0.96\}$. At the mean redshift of the 4MOST+LSST sample, we expect $f=0.777$ and adopt this as fiducial value. For the clustering bias, we adopt a value consistent with the estimate for the LOWZ sample of the BOSS survey \citep{Singh}: $b_g=1.77\pm 0.04$. Following that work, we take $\tilde{C}_1\simeq 0.026$ as our fiducial alignment amplitude, with a $10\%$ uncertainty.  We also use $\eta=4$ and $\chi=0.06$ based on \citet{Hirata09}, which implies $(\chi\eta)_{\rm eff}=0.24$. We take this value as fixed. In practice, this will be dependent on the selection function of the survey and should be estimated with dedicated simulations. 

\section{Results}
\label{sec:results}

We present the expected constraints on $b_g$, $f$ and $\tilde{C_1}$ in this section. We also present results for the figure of merit FoM in the $(b_g,f)$ parameter space. Our main results are condensed in Table \ref{tab:results} and we will discuss them throughout this section. We discuss first the results for $k_{\rm max}=0.3$ Mpc$^{-1}$, but we vary this choice towards the end of the section.

\begin{table*}
    \centering
    \begin{tabular}{l|c|c|c|c}
    \hline
    \hline
        {\rm Study case} & $\sigma_{b_g}/b_g$ & $\sigma_{f}/f$ & $\sigma_{\tilde{C}_1}/\tilde{C}_1$ & Figure of merit  \\
        \hline\hline
        $k_{\rm max}=0.1$ Mpc$^{-1}$ & & & &\\
        \hline
        $\{P_{gg}^{(0)},P_{gg}^{(2)}\}$ no IA & $3.1 \times 10^{-2}$ & $1.8 \times 10^{-2}$ & - & 8300 \\
        $\{P_{gg}^{(0)},P_{gg}^{(2)}\}$, $\tilde{C_1}$ prior & $3.1 \times 10^{-2}$ & $1.8 \times 10^{-2}$ & $1.1 \times 10^{-1}$& 8200 \\
        $\{P_{gg}^{(0)},P_{gg}^{(2)},P_{gE}^{(0)}\}$ & $3.1 \times 10^{-2}$ & $1.8 \times 10^{-2}$ & $2.6 \times 10^{-2}$&  8300\\
        $\{P_{gg}^{(0)},P_{gg}^{(2)},P_{EE}^{(0)}\}$ & $3.1 \times 10^{-2}$ & $1.8 \times 10^{-2}$ & $1.2 \times 10^{-1}$& 8200 \\
         \hline
        $k_{\rm max}=0.2$ Mpc$^{-1}$ & & & & \\
        \hline
        $\{P_{gg}^{(0)},P_{gg}^{(2)}\}$ no IA & $  1.3\times 10^{-3}$& $ 7.2\times 10^{-3}$& - & $49\times 10^3$\\ 
        $\{P_{gg}^{(0)},P_{gg}^{(2)}\}$, $\tilde{C_1}$ prior & $1.4\times 10^{-3}$ & $7.6\times 10^{-3}$ & $ 1.1\times 10^{-1}$ & $46\times 10^3$\\ 
        $\{P_{gg}^{(0)},P_{gg}^{(2)},P_{gE}^{(0)}\}$ & $1.3\times 10^{-3}$& $7.2\times 10^{-3}$ & $1.6\times 10^{-2}$& $49\times 10^3$\\
        $\{P_{gg}^{(0)},P_{gg}^{(2)},P_{EE}^{(0)}\}$ & $1.4\times 10^{-3}$ & $7.6\times 10^{-3}$& $1.0\times 10^{-1}$& $ 48\times 10^3$\\
        \hline
        $k_{\rm max}=0.3$ Mpc$^{-1}$ & & & & \\
        \hline
        $\{P_{gg}^{(0)},P_{gg}^{(2)}\}$ no IA & $8.7\times 10^{-4}$& $4.7\times 10^{-3}$& - & $110\times 10^3$\\ 
        $\{P_{gg}^{(0)},P_{gg}^{(2)}\}$, $\tilde{C_1}$ prior & $9.8\times 10^{-4}$ & $5.3\times 10^{-3}$ & $1.1 \times 10^{-1}$ & $99 \times 10^3$\\
        $\{P_{gg}^{(0)},P_{gg}^{(2)},P_{gE}^{(0)}\}$ & $8.8\times 10^{-4}$& $4.7\times 10^{-3}$ & $1.3\times 10^{-2}$& $110\times 10^3$\\
        $\{P_{gg}^{(0)},P_{gg}^{(2)},P_{EE}^{(0)}\}$ & $9.6\times 10^{-4}$ & $5.2\times 10^{-3}$ & $9.5\times 10^{-2}$ & $100 \times 10^3$ \\
        \hline
    \end{tabular}
    \centering
    \caption{Estimated $1\sigma$ uncertainties in the parameters of interest, marginalised over all other parameters, and figure of merit in the $(b_g,f)$ parameter space for different choices of $k_{\rm max}$. The first line corresponds to an idealised case of no alignments in the Universe, nor in the model. The second line includes a prior on $\tilde{C_1}$ from \citet{Singh}. The subsequent lines combine the clustering multipoles with different alignment observables.}
    \label{tab:results}
\end{table*}

\begin{figure}
    \centering
    \includegraphics[width=0.45\textwidth]{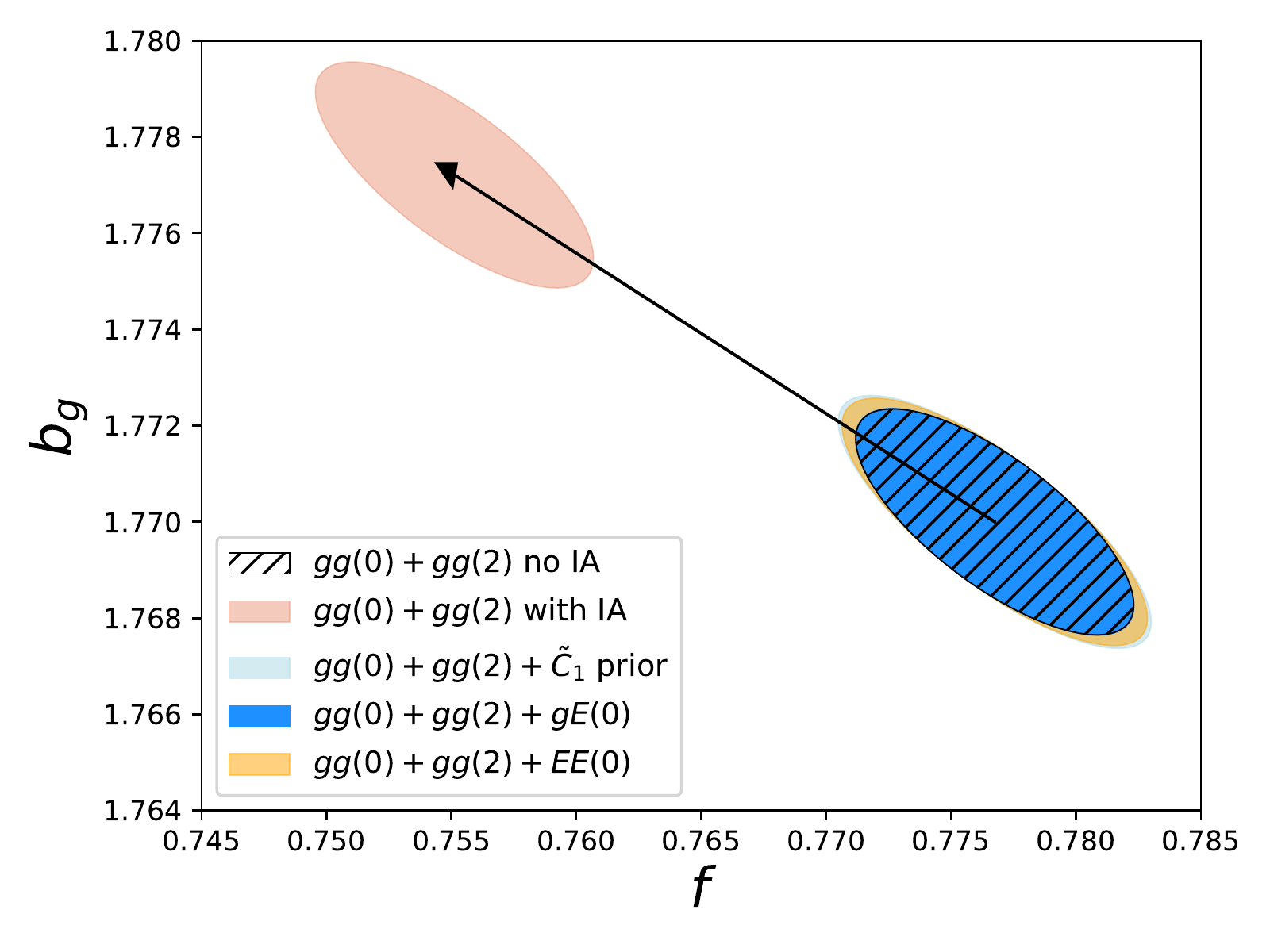}
    \caption{Expected parameter constraints ($68\%$ confidence level) on $b_g$ and $f$ from different probe combinations when $k_{\rm max}=0.3$ Mpc$^{-1}$. The hatched contour indicates an idealised case of no alignments in the Universe, and none in the model. The pink ellipse is shifted to the inferred $(b_g-A/3,f+A)$ when alignments are missing in the model (notice $A$ is negative). Other ellipses consider cases where alignments are included in the model. The light blue contour considers an analysis of the clustering monopole and quadrupole with a prior for $\tilde{C_1}$. The dark blue contour includes the position-shape monopole. The orange contour includes the shape-shape monopole instead of position-shape.}
    \label{fig:bgf}
\end{figure}

We first calculate the potential constraints on $b_g$ and $f$ from the clustering monopole $P_{gg}^{(0)}$ and quadrupole $P_{gg}^{(2)}$ if intrinsic alignments were not present in the Universe nor in the model. For this idealised case, we find that the expected $1\sigma$ ($68\%$ confidence level) uncertainties in the parameters are of the order of $0.087\%$ and $0.47\%$ for $b_g$ and $f$, respectively.
These are, of course, correlated, which can be seen in the hatched ellipse in Figure \ref{fig:bgf}. Such scenario provides us with a benchmark against which to compare the values of the FoMs presented in Table \ref{tab:results}. For $k_{\rm max}=0.3\,h^{-1}\,{\rm Mpc}$, the FoM is $110 \times 10^3$.

Secondly, we use Eqs. (\ref{eq:bias }) and (\ref{eq:biasvec}) to estimate, as a sanity check, the bias in $b_g$ and $f$ from neglecting the selection effect induced by intrinsic alignments in the clustering signal. We confirm that neglecting intrinsic alignments in the model, while they are present in the data, results in the estimated bias being $\hat{b}_g\simeq b_g-A/3$, and the growth rate being $\hat{f}\simeq f+A$ (where $A$ is a negative number), as we anticipated in Section \ref{sec:gg}. This is shown by the pink ellipse in Figure \ref{fig:bgf}.

\begin{figure}
    \centering
    \includegraphics[width=0.45\textwidth]{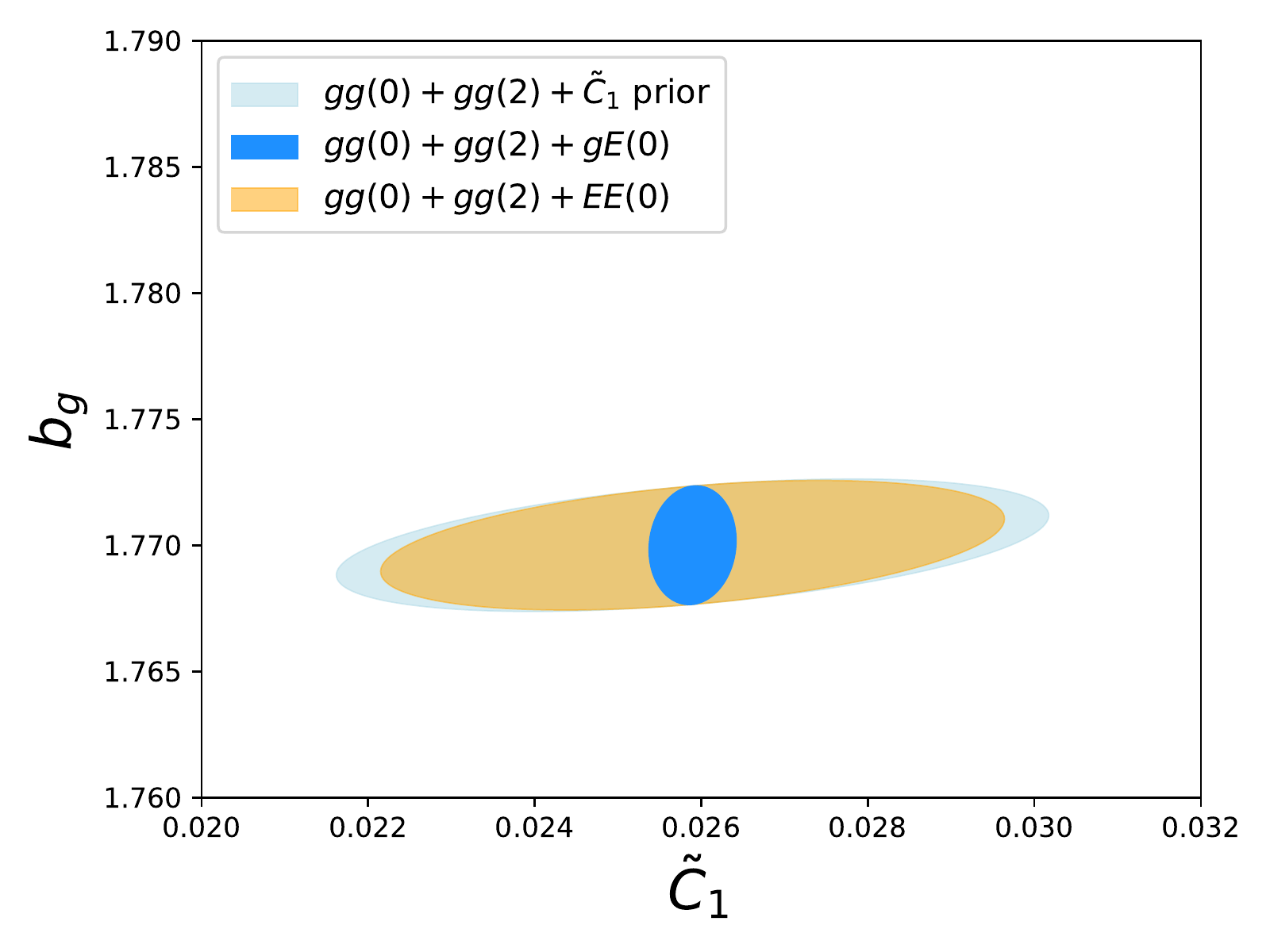}
    \caption{Expected parameter constraints ($68\%$ confidence level) on $b_g$ and $\tilde{C_1}$ from different probe combinations when $k_{\rm max}=0.3$ Mpc$^{-1}$. In all cases, alignments are included in the model. The light blue contour considers an analysis of the clustering monopole and quadrupole with a prior for $\tilde{C_1}$ coming from observations \citep{Singh}. The dark blue contour includes the position-shape monopole. The orange contour includes the shape-shape monopole instead of position-shape.}
    \label{fig:bgc1}
\end{figure}

Next, we include the selection effects induced by intrinsic alignments in the clustering measurements, and combine the clustering monopole $P_{gg}^{(0)}$ and quadrupole $P_{gg}^{(2)}$ once again to obtain constraints on the bias and the growth rate. When including a prior on $\tilde{C_1}$ from \citet{Singh}, the expected marginalised $1\sigma$ uncertainties in this case are $0.098\%$ and $0.53\%$ for $b_g$ and $f$, respectively. This case is shown in the light blue contour in Figure \ref{fig:bgf}. Figure \ref{fig:bgc1} also shows the degeneracy between $b_g$ and $\tilde{C_1}$. The size of the light blue contour is largely driven by the $\tilde{C_1}$ prior ($\sim 10\%$). Importantly, this scenario is not completely realistic because the prior on $\tilde{C_1}$ comes from measuring projected correlation functions of galaxy alignments and galaxy clustering in \citet{Singh} and these would not be independent of the observables we consider. Nevertheless, one might have access to an external prior from other observables or from simulations, so it is still a useful scenario to consider. The inclusion of intrinsic alignments degrades the figure of merit, reducing it to $99 \times 10^3$.

There is more to be gained from the joint modelling of alignments and clustering multipoles. If we remove the prior and instead consider a data vector comprised of $\{P_{gg}^{(0)},P_{gg}^{(2)},P_{gE}^{(0)}\}$, the marginalised $1\sigma$ uncertainties are comparable to those in the idealised case (black hatched ellipse in Figure \ref{fig:bgf}). Similarly, for our fiducial noise model, we see that this scenario recovers the same figure of merit as the no alignment scenario: $110 \times 10^3$. The reason is that depending on the number density of galaxies, $P_{gE}^{(0)}$ can add information on $b_g$ and $f$. We verified this by varying the number density and confirming that lower values decrease the figure of merit compared to the idealised case, while higher values correspondingly increase it. 

Finally, we assess the constraining power in $P_{EE}^{(0)}$ by considering the combination $\{P_{gg}^{(0)},P_{gg}^{(2)},P_{EE}^{(0)}\}$. Due to the increased noise in $P_{EE}^{(0)}$ compared to $P_{gE}^{(0)}$, the constrains here are not as good as in the previous case: $0.096\%$ and $0.52\%$ for $b_g$ and $f$, respectively. The figure of merit is also reduced in this case ($100 \times 10^3$) compared to the target scenario of no alignments. Similarly, we find that the combination $\{P_{gg}^{(0)},P_{gg}^{(2)},P_{gE}^{(0)},P_{EE}^{(0)}\}$ does not increase the constraining power significantly over the case where the shape-shape monopole is excluded. This is to be expected given that $P_{gE}^{(0)}$ is not affected by the noise contribution coming from the intrinsic dispersion of galaxy ellipticities, and has thus increased constraining power over $P_{EE}^{(0)}$. This fact is also, in fact, convenient from the point of view of lensing contamination. It is easy to isolate the alignment signal in $P_{gE}^{(0)}$ through restrictions in the relative distance between the galaxies. In the case of $P_{EE}^{(0)}$, restricting the redshift range between galaxy pairs cannot remove the lensing contamination. Galaxies at the same redshift would be indeed lensed by the same foreground structures and their shapes would be correlated. In this case, both lensing and alignments would have to be modelled jointly.


\subsection{Impact of parameter choices}

\begin{figure}
    \centering
    \includegraphics[width=0.45\textwidth]{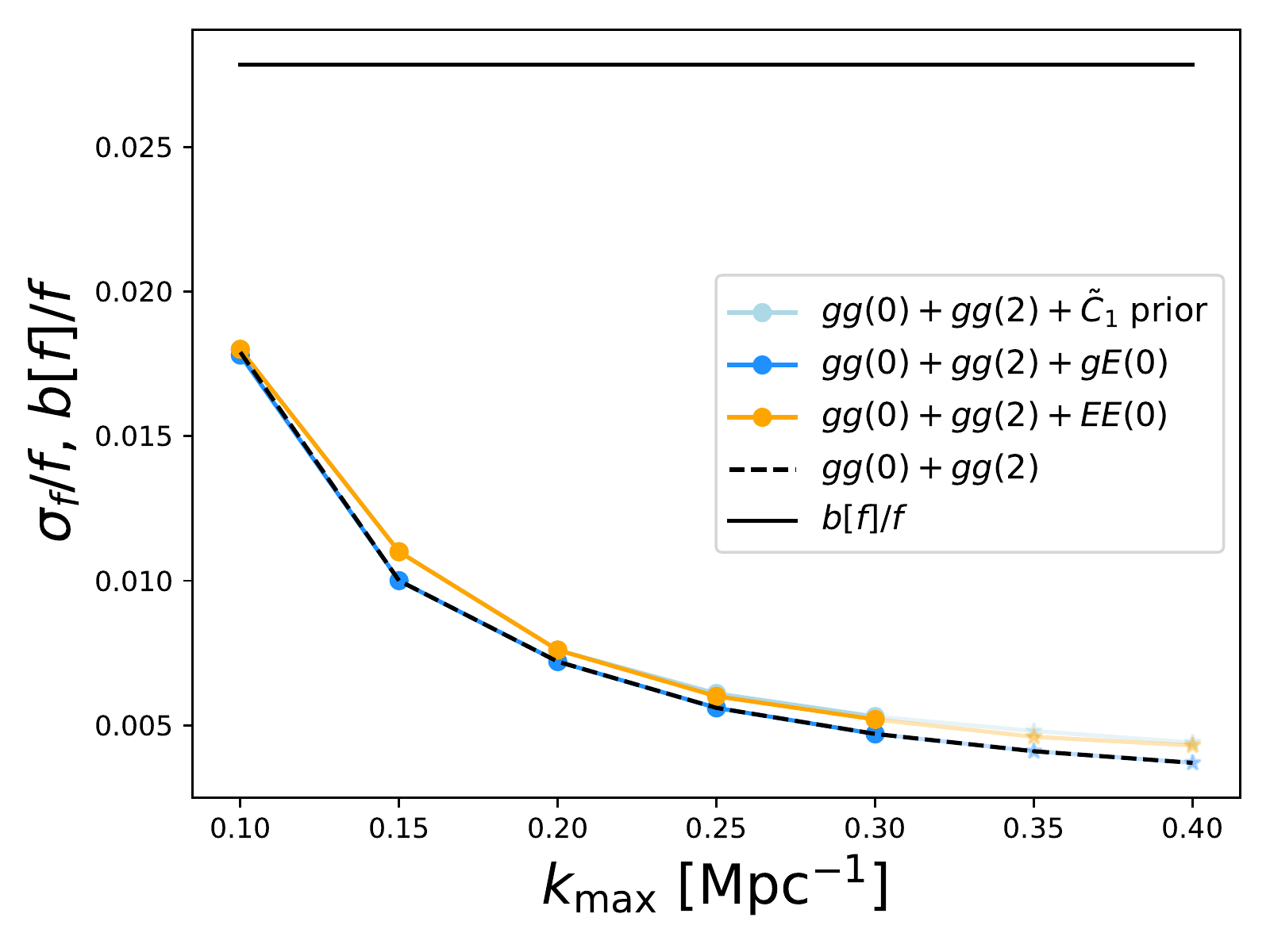}
    \caption{Expected fractional uncertainty in $f$ (at the 68\% confidence level) for various choices of $k_{\rm max}$ between 0.1 Mpc$^{-1}$ and 0.4 Mpc$^{-1}$. The marginalised uncertainty is always significantly smaller than the expected bias if alignments are neglected. Ths is exacerbated for increasing $k_{\rm max}$ due to the enhanced constraining power brought in by the inclusion of more modes.}
    \label{fig:kmax}
\end{figure}

The impact of varying $k_{\rm max}$ is significant in the analysis, given that the number of modes increases as $\propto k^3$. For a conservative choice of $k_{\rm max}=0.1$ Mpc$^{-1}$, we find that intrinsic alignments have little impact on the $(b_g,f)$ figure of merit. The bias in the estimated parameters is also significantly reduced and it overlaps at $<1\sigma$ with the true parameters. Table \ref{tab:results} shows also an intermediate case with $k_{\rm max}=0.2$ Mpc$^{-1}$. Moreover, Figure \ref{fig:kmax} shows the fractional uncertainty in $f$ as a function of $k_{\rm max}$ for our different analysis scenarios. In all cases, the uncertainty is smaller than the expected bias. It also decreases with $k_{\rm max}$, as expected due to the inclusion of more modes in the analysis. Results should be interpreted with caution, since our methodology relies on a fully linear model. At the moment, this is the only model available that can at the same time account for IA and RSD.

We proceed to keep $k_{\rm max}=0.3$ Mpc$^{-1}$ constant and explore the sensitivity of our results to the choice of other parameters in our analysis. Varying $\Omega_{\rm cdm}$ has an impact on the derived growth rate $f$ and the expected uncertainties only for the case where a prior on $\tilde{C}_1$ is considered. Mainly, if the prior remains constant, a decrease in $f$ results in the data having more difficulties in constraining it. Changes in $\Omega_{\rm cdm}$ of the order of $<10\%$ result in a 
few percent changes on the fractional uncertainty on $f$. When clustering and alignments are combined, this limitation is lifted since the alignment amplitude is constrained jointly with $f$.

On the other hand, increasing the prior information on $\tilde{C}_1$ improves the expected constraints on $f$. But to achieve the same constraining power as when $P_{\rm gE}(0)$ is included, we need to improve the knowledge of the prior by a factor of $\sim 5$. This result highlights the advantage of combining probes versus simply relying on a prior for $\tilde{C}_1$.

Finally, we have looked at the impact of $(\eta\chi)_{\rm eff}$ on our results. Because the connection between $\tilde{C}_1$ and $A$ is deterministic, we can state that the expected bias in $f$ if alignments are neglected grows linearly with $(\eta\chi)_{\rm eff}$. We have also estimated the value of $(\eta\chi)_{\rm eff}$ at which the expected bias and uncertainty in $f$ are the same. This occurs when $(\eta\chi)_{\rm eff}\simeq 0.05$, which is a factor of $5$ below our currently adopted value. Clustering multipoles are thus very sensitive to the selection effects induced by alignments: to the point that only a factor $5$ decrease in our assumed $(\eta\chi)_{\rm eff}$ could make the bias comparable to the statistical uncertainty.

\section{Conclusions}
\label{sec:conclusions}

Intrinsic alignments can cause several $\sigma$ biased constraints on the bias of galaxies and the growth rate depending on the maximum wave number of the analysis. Conservative analyses up to $k_{\rm max}=0.1$ Mpc$^{-1}$ will not suffer an impact, but extending the analysis to higher scales will result in a significant bias compared to the measurement uncertainties. 

Adding a prior on the alignment amplitude can mitigate the impact of alignments. A better alternative is to add the position-intrinsic shape monopole to the data vector. This serves not only to mitigate the impact of alignments, but to also inform parameter constraints depending on the number density of galaxies. The sensitivity to the underlying cosmology is smaller when $gE(0)$ is added, compared to the case of using a prior on $\tilde{C}_1$.

Some assumptions made in this analysis should be explored further to fit to the needs of spectroscopic experiments. In taking our fiducial $\tilde{C}_1$ value from \citep{Singh}, we are extrapolating their measurements to a higher mean redshift, neglecting a potential luminosity and redshift evolution of alignments and neglecting small variations in the underlying cosmological model between that work and ours. Among these, the strongest assumption is the absence of redshift and luminosity evolution in $\tilde{C}_1$. However, such behaviour is difficult to predict, since it also depends on the fraction of satellites in the galaxy sample \citep{Fortuna}. We thus decided not to include it at this stage.

In the future, we plan to expand our work to make the analysis and model set-up more realistic in the quasilinear and nonlinear regimes. This can be done, for example, by employing perturbative models for galaxy alignments and galaxy clustering \citep{Blazek,Vlah,Zhang22} or by using a halo model approach \citep{Fortuna}. More realistic estimates of contamination would have to be tailored to specific surveys by means of mock catalogues and by sampling the likelihood of the cosmological and nuisance parameter space. Although this work has focused on spectroscopic samples, photometric ones might also be affected by the selection effects presented here. We expect this would be at a lower level, since redshift uncertainties coming from photometric data complicate potential constraints on the growth rate.

Degeneracies between $b_g$ and $f$ can be broken by the addition of higher order statistics, for example \citep{Verde02,GilMarinBispec}. Hence, it would also be interesting to consider the impact of alignment contamination to such probe combinations and whether higher order clustering or alignment statistics could also mitigate this scenario or even add information \citep{Pyne,Agarwal21}. We should also note that in this work we chose not to marginalise over the amplitude of fluctuations in the density field today (typically parameterised by $\sigma_8$ or $A_s$). This choice could be relaxed in future work, and the degeneracy between $b_g$, $f$ and $\sigma_8$ broken by combining with other probes, such as cosmic shear.

\section*{Acknowledgements}

This publication is part of the project ``A rising tide: Galaxy intrinsic alignments as a new probe of cosmology and galaxy evolution'' (with project number VI.Vidi.203.011) of the Talent programme Vidi which is (partly) financed by the Dutch Research Council (NWO). This work is also part of the Delta ITP consortium, a program of the Netherlands Organisation for Scientific Research (NWO) that is funded by the Dutch Ministry of Education, Culture and Science (OCW). We thank Harry Johnston for comments that helped improve this manuscript.

\section*{Data Availability}

No new data was used in this project. The software tool for all theoretical predictions ({\tt CCL v2.3}) is due to the Dark Energy Science Collaboration of the Vera C. Rubin Observatory and publicly available at \url{https://github.com/lsstDESC/CCL}. The matter power spectrum was computed relying on the {\tt CLASS} software \citep{CLASS}.



\bibliographystyle{mnras}
\bibliography{biblioIA} 

\begin{thebibliography}{}
\makeatletter
\relax
\def\mn@urlcharsother{\let\do\@makeother \do\$\do\&\do\#\do\^\do\_\do\%\do\~}
\def\mn@doi{\begingroup\mn@urlcharsother \@ifnextchar [ {\mn@doi@}
  {\mn@doi@[]}}
\def\mn@doi@[#1]#2{\def\@tempa{#1}\ifx\@tempa\@empty \href
  {http://dx.doi.org/#2} {doi:#2}\else \href {http://dx.doi.org/#2} {#1}\fi
  \endgroup}
\def\mn@eprint#1#2{\mn@eprint@#1:#2::\@nil}
\def\mn@eprint@arXiv#1{\href {http://arxiv.org/abs/#1} {{\tt arXiv:#1}}}
\def\mn@eprint@dblp#1{\href {http://dblp.uni-trier.de/rec/bibtex/#1.xml}
  {dblp:#1}}
\def\mn@eprint@#1:#2:#3:#4\@nil{\def\@tempa {#1}\def\@tempb {#2}\def\@tempc
  {#3}\ifx \@tempc \@empty \let \@tempc \@tempb \let \@tempb \@tempa \fi \ifx
  \@tempb \@empty \def\@tempb {arXiv}\fi \@ifundefined
  {mn@eprint@\@tempb}{\@tempb:\@tempc}{\expandafter \expandafter \csname
  mn@eprint@\@tempb\endcsname \expandafter{\@tempc}}}

\bibitem[\protect\citeauthoryear{{Agarwal}, {Desjacques}, {Jeong}  \&
  {Schmidt}}{{Agarwal} et~al.}{2021}]{Agarwal21}
{Agarwal} N.,  {Desjacques} V.,  {Jeong} D.,   {Schmidt} F.,  2021, \mn@doi
  [\jcap] {10.1088/1475-7516/2021/03/021}, \href
  {https://ui.adsabs.harvard.edu/abs/2021JCAP...03..021A} {2021, 021}

\bibitem[\protect\citeauthoryear{{Alam} et~al.,}{{Alam} et~al.}{2017}]{Alam17}
{Alam} S.,  et~al., 2017, \mn@doi [\mnras] {10.1093/mnras/stx721}, \href
  {https://ui.adsabs.harvard.edu/abs/2017MNRAS.470.2617A} {470, 2617}

\bibitem[\protect\citeauthoryear{{Alam} et~al.,}{{Alam} et~al.}{2021}]{Alam21}
{Alam} S.,  et~al., 2021, \mn@doi [\prd] {10.1103/PhysRevD.103.083533}, \href
  {https://ui.adsabs.harvard.edu/abs/2021PhRvD.103h3533A} {103, 083533}

\bibitem[\protect\citeauthoryear{{Albrecht} et~al.,}{{Albrecht}
  et~al.}{2006}]{DETF}
{Albrecht} A.,  et~al., 2006, arXiv e-prints, \href
  {https://ui.adsabs.harvard.edu/abs/2006astro.ph..9591A} {pp
  astro--ph/0609591}

\bibitem[\protect\citeauthoryear{{Amara} \& {R{\'e}fr{\'e}gier}}{{Amara} \&
  {R{\'e}fr{\'e}gier}}{2008}]{Amara}
{Amara} A.,  {R{\'e}fr{\'e}gier} A.,  2008, \mn@doi [\mnras]
  {10.1111/j.1365-2966.2008.13880.x}, \href
  {https://ui.adsabs.harvard.edu/abs/2008MNRAS.391..228A} {391, 228}

\bibitem[\protect\citeauthoryear{{Bautista} et~al.,}{{Bautista}
  et~al.}{2021}]{Bautista2021}
{Bautista} J.~E.,  et~al., 2021, \mn@doi [\mnras] {10.1093/mnras/staa2800},
  \href {https://ui.adsabs.harvard.edu/abs/2021MNRAS.500..736B} {500, 736}

\bibitem[\protect\citeauthoryear{{Bernstein}}{{Bernstein}}{2009}]{Bernstein09}
{Bernstein} G.~M.,  2009, \mn@doi [\apj] {10.1088/0004-637X/695/1/652}, \href
  {https://ui.adsabs.harvard.edu/abs/2009ApJ...695..652B} {695, 652}

\bibitem[\protect\citeauthoryear{{Beutler} et~al.,}{{Beutler}
  et~al.}{2012}]{Beutler12}
{Beutler} F.,  et~al., 2012, \mn@doi [\mnras]
  {10.1111/j.1365-2966.2012.21136.x}, \href
  {https://ui.adsabs.harvard.edu/abs/2012MNRAS.423.3430B} {423, 3430}

\bibitem[\protect\citeauthoryear{{Blake}, {Carter}  \& {Koda}}{{Blake}
  et~al.}{2018}]{Blake18}
{Blake} C.,  {Carter} P.,   {Koda} J.,  2018, \mn@doi [\mnras]
  {10.1093/mnras/sty1814}, \href
  {https://ui.adsabs.harvard.edu/abs/2018MNRAS.479.5168B} {479, 5168}

\bibitem[\protect\citeauthoryear{{Blas}, {Lesgourgues}  \& {Tram}}{{Blas}
  et~al.}{2011}]{CLASS}
{Blas} D.,  {Lesgourgues} J.,   {Tram} T.,  2011, \mn@doi [\jcap]
  {10.1088/1475-7516/2011/07/034}, \href
  {https://ui.adsabs.harvard.edu/abs/2011JCAP...07..034B} {2011, 034}

\bibitem[\protect\citeauthoryear{{Blazek}, {McQuinn}  \& {Seljak}}{{Blazek}
  et~al.}{2011}]{Blazek11}
{Blazek} J.,  {McQuinn} M.,   {Seljak} U.,  2011, \mn@doi [\jcap]
  {10.1088/1475-7516/2011/05/010}, \href
  {https://ui.adsabs.harvard.edu/abs/2011JCAP...05..010B} {2011, 010}

\bibitem[\protect\citeauthoryear{{Blazek}, {MacCrann}, {Troxel}  \&
  {Fang}}{{Blazek} et~al.}{2019}]{Blazek}
{Blazek} J.~A.,  {MacCrann} N.,  {Troxel} M.~A.,   {Fang} X.,  2019, \mn@doi
  [\prd] {10.1103/PhysRevD.100.103506}, \href
  {https://ui.adsabs.harvard.edu/abs/2019PhRvD.100j3506B} {100, 103506}

\bibitem[\protect\citeauthoryear{{Catelan}, {Kamionkowski}  \&
  {Blandford}}{{Catelan} et~al.}{2001}]{Catelan01}
{Catelan} P.,  {Kamionkowski} M.,   {Blandford} R.~D.,  2001, \mn@doi [\mnras]
  {10.1046/j.1365-8711.2001.04105.x}, \href
  {https://ui.adsabs.harvard.edu/abs/2001MNRAS.320L...7C} {320, L7}

\bibitem[\protect\citeauthoryear{{Chapman} et~al.,}{{Chapman}
  et~al.}{2021}]{Chapman21}
{Chapman} M.~J.,  et~al., 2021, arXiv e-prints, \href
  {https://ui.adsabs.harvard.edu/abs/2021arXiv210614961C} {p. arXiv:2106.14961}

\bibitem[\protect\citeauthoryear{{Chen}, {Vlah}  \& {White}}{{Chen}
  et~al.}{2022}]{Chen22}
{Chen} S.-F.,  {Vlah} Z.,   {White} M.,  2022, \mn@doi [\jcap]
  {10.1088/1475-7516/2022/02/008}, \href
  {https://ui.adsabs.harvard.edu/abs/2022JCAP...02..008C} {2022, 008}

\bibitem[\protect\citeauthoryear{{Chisari} et~al.,}{{Chisari}
  et~al.}{2019}]{Chisari19}
{Chisari} N.~E.,  et~al., 2019, \mn@doi [\apjs] {10.3847/1538-4365/ab1658},
  \href {https://ui.adsabs.harvard.edu/abs/2019ApJS..242....2C} {242, 2}

\bibitem[\protect\citeauthoryear{{Dawson} et~al.,}{{Dawson}
  et~al.}{2013}]{Dawson13}
{Dawson} K.~S.,  et~al., 2013, \mn@doi [\aj] {10.1088/0004-6256/145/1/10},
  \href {https://ui.adsabs.harvard.edu/abs/2013AJ....145...10D} {145, 10}

\bibitem[\protect\citeauthoryear{{Djorgovski} \& {Davis}}{{Djorgovski} \&
  {Davis}}{1987}]{Djorgovski87}
{Djorgovski} S.,  {Davis} M.,  1987, \mn@doi [\apj] {10.1086/164948}, \href
  {https://ui.adsabs.harvard.edu/abs/1987ApJ...313...59D} {313, 59}

\bibitem[\protect\citeauthoryear{{Fortuna}, {Hoekstra}, {Joachimi}, {Johnston},
  {Chisari}, {Georgiou}  \& {Mahony}}{{Fortuna} et~al.}{2021a}]{Fortuna}
{Fortuna} M.~C.,  {Hoekstra} H.,  {Joachimi} B.,  {Johnston} H.,  {Chisari}
  N.~E.,  {Georgiou} C.,   {Mahony} C.,  2021a, \mn@doi [\mnras]
  {10.1093/mnras/staa3802}, \href
  {https://ui.adsabs.harvard.edu/abs/2021MNRAS.501.2983F} {501, 2983}

\bibitem[\protect\citeauthoryear{{Fortuna} et~al.,}{{Fortuna}
  et~al.}{2021b}]{Fortuna21}
{Fortuna} M.~C.,  et~al., 2021b, \mn@doi [\aap] {10.1051/0004-6361/202140706},
  \href {https://ui.adsabs.harvard.edu/abs/2021A&A...654A..76F} {654, A76}

\bibitem[\protect\citeauthoryear{{Gil-Mar{\'\i}n}, {Percival}, {Verde},
  {Brownstein}, {Chuang}, {Kitaura}, {Rodr{\'\i}guez-Torres}  \&
  {Olmstead}}{{Gil-Mar{\'\i}n} et~al.}{2017}]{GilMarinBispec}
{Gil-Mar{\'\i}n} H.,  {Percival} W.~J.,  {Verde} L.,  {Brownstein} J.~R.,
  {Chuang} C.-H.,  {Kitaura} F.-S.,  {Rodr{\'\i}guez-Torres} S.~A.,
  {Olmstead} M.~D.,  2017, \mn@doi [\mnras] {10.1093/mnras/stw2679}, \href
  {https://ui.adsabs.harvard.edu/abs/2017MNRAS.465.1757G} {465, 1757}

\bibitem[\protect\citeauthoryear{{Hirata}}{{Hirata}}{2009}]{Hirata09}
{Hirata} C.~M.,  2009, \mn@doi [\mnras] {10.1111/j.1365-2966.2009.15353.x},
  \href {https://ui.adsabs.harvard.edu/abs/2009MNRAS.399.1074H} {399, 1074}

\bibitem[\protect\citeauthoryear{{Hirata} \& {Seljak}}{{Hirata} \&
  {Seljak}}{2004}]{Hirata04}
{Hirata} C.~M.,  {Seljak} U.,  2004, \mn@doi [\prd]
  {10.1103/PhysRevD.70.063526}, \href
  {https://ui.adsabs.harvard.edu/abs/2004PhRvD..70f3526H} {70, 063526}

\bibitem[\protect\citeauthoryear{{Hirata}, {Mandelbaum}, {Ishak}, {Seljak},
  {Nichol}, {Pimbblet}, {Ross}  \& {Wake}}{{Hirata} et~al.}{2007}]{Hirata07}
{Hirata} C.~M.,  {Mandelbaum} R.,  {Ishak} M.,  {Seljak} U.,  {Nichol} R.,
  {Pimbblet} K.~A.,  {Ross} N.~P.,   {Wake} D.,  2007, \mn@doi [\mnras]
  {10.1111/j.1365-2966.2007.12312.x}, \href
  {https://ui.adsabs.harvard.edu/abs/2007MNRAS.381.1197H} {381, 1197}

\bibitem[\protect\citeauthoryear{{Hui} \& {Zhang}}{{Hui} \&
  {Zhang}}{2002}]{Hui}
{Hui} L.,  {Zhang} J.,  2002, arXiv e-prints, \href
  {https://ui.adsabs.harvard.edu/abs/2002astro.ph..5512H} {pp
  astro--ph/0205512}

\bibitem[\protect\citeauthoryear{{Ivanov}}{{Ivanov}}{2021}]{Ivanov21}
{Ivanov} M.~M.,  2021, \mn@doi [\prd] {10.1103/PhysRevD.104.103514}, \href
  {https://ui.adsabs.harvard.edu/abs/2021PhRvD.104j3514I} {104, 103514}

\bibitem[\protect\citeauthoryear{{Ivezi{\'c}} et~al.}{{Ivezi{\'c}}
  et~al.}{2019}]{LSST}
{Ivezi{\'c}} {\v{Z}}.,  et~al., 2019, \mn@doi [\apj]
  {10.3847/1538-4357/ab042c}, \href
  {https://ui.adsabs.harvard.edu/abs/2019ApJ...873..111I} {873, 111}

\bibitem[\protect\citeauthoryear{{Jain} \& {Zhang}}{{Jain} \&
  {Zhang}}{2008}]{Jain08}
{Jain} B.,  {Zhang} P.,  2008, \mn@doi [\prd] {10.1103/PhysRevD.78.063503},
  \href {https://ui.adsabs.harvard.edu/abs/2008PhRvD..78f3503J} {78, 063503}

\bibitem[\protect\citeauthoryear{{Joachimi}, {Mandelbaum}, {Abdalla}  \&
  {Bridle}}{{Joachimi} et~al.}{2011}]{Joachimi}
{Joachimi} B.,  {Mandelbaum} R.,  {Abdalla} F.~B.,   {Bridle} S.~L.,  2011,
  \mn@doi [\aap] {10.1051/0004-6361/201015621}, \href
  {https://ui.adsabs.harvard.edu/abs/2011A&A...527A..26J} {527, A26}

\bibitem[\protect\citeauthoryear{{Johnston} et~al.,}{{Johnston}
  et~al.}{2019}]{Johnston}
{Johnston} H.,  et~al., 2019, \mn@doi [\aap] {10.1051/0004-6361/201834714},
  \href {https://ui.adsabs.harvard.edu/abs/2019A&A...624A..30J} {624, A30}

\bibitem[\protect\citeauthoryear{{Kaiser}}{{Kaiser}}{1987}]{Kaiser87}
{Kaiser} N.,  1987, \mn@doi [\mnras] {10.1093/mnras/227.1.1}, \href
  {https://ui.adsabs.harvard.edu/abs/1987MNRAS.227....1K} {227, 1}

\bibitem[\protect\citeauthoryear{{Kamionkowski}, {Kosowsky}  \&
  {Stebbins}}{{Kamionkowski} et~al.}{1997}]{Kamionkowski}
{Kamionkowski} M.,  {Kosowsky} A.,   {Stebbins} A.,  1997, \mn@doi [\prd]
  {10.1103/PhysRevD.55.7368}, \href
  {https://ui.adsabs.harvard.edu/abs/1997PhRvD..55.7368K} {55, 7368}

\bibitem[\protect\citeauthoryear{{Kobayashi}, {Nishimichi}, {Takada}  \&
  {Miyatake}}{{Kobayashi} et~al.}{2022}]{Kobayashi22}
{Kobayashi} Y.,  {Nishimichi} T.,  {Takada} M.,   {Miyatake} H.,  2022, \mn@doi
  [\prd] {10.1103/PhysRevD.105.083517}, \href
  {https://ui.adsabs.harvard.edu/abs/2022PhRvD.105h3517K} {105, 083517}

\bibitem[\protect\citeauthoryear{{Lange}, {Hearin}, {Leauthaud}, {van den
  Bosch}, {Guo}  \& {DeRose}}{{Lange} et~al.}{2022}]{Lange22}
{Lange} J.~U.,  {Hearin} A.~P.,  {Leauthaud} A.,  {van den Bosch} F.~C.,  {Guo}
  H.,   {DeRose} J.,  2022, \mn@doi [\mnras] {10.1093/mnras/stab3111}, \href
  {https://ui.adsabs.harvard.edu/abs/2022MNRAS.509.1779L} {509, 1779}

\bibitem[\protect\citeauthoryear{{Mandelbaum}, {Hirata}, {Ishak}  \&
  {Seljak}}{{Mandelbaum} et~al.}{2006}]{Mandelbaum06}
{Mandelbaum} R.,  {Hirata} C.~M.,  {Ishak} M.,   {Seljak} U.,  2006, in
  American Astronomical Society Meeting Abstracts. p. 77.27

\bibitem[\protect\citeauthoryear{{Martens}, {Hirata}, {Ross}  \&
  {Fang}}{{Martens} et~al.}{2018}]{Martens18}
{Martens} D.,  {Hirata} C.~M.,  {Ross} A.~J.,   {Fang} X.,  2018, \mn@doi
  [\mnras] {10.1093/mnras/sty1100}, \href
  {https://ui.adsabs.harvard.edu/abs/2018MNRAS.478..711M} {478, 711}

\bibitem[\protect\citeauthoryear{{Percival} \& {White}}{{Percival} \&
  {White}}{2009}]{Percival09}
{Percival} W.~J.,  {White} M.,  2009, \mn@doi [\mnras]
  {10.1111/j.1365-2966.2008.14211.x}, \href
  {https://ui.adsabs.harvard.edu/abs/2009MNRAS.393..297P} {393, 297}

\bibitem[\protect\citeauthoryear{{Philcox} \& {Ivanov}}{{Philcox} \&
  {Ivanov}}{2022}]{Philcox22}
{Philcox} O. H.~E.,  {Ivanov} M.~M.,  2022, \mn@doi [\prd]
  {10.1103/PhysRevD.105.043517}, \href
  {https://ui.adsabs.harvard.edu/abs/2022PhRvD.105d3517P} {105, 043517}

\bibitem[\protect\citeauthoryear{{Planck Collaboration} et~al.,}{{Planck
  Collaboration} et~al.}{2020}]{Planck}
{Planck Collaboration} et~al., 2020, \mn@doi [\aap]
  {10.1051/0004-6361/201833910}, \href
  {https://ui.adsabs.harvard.edu/abs/2020A&A...641A...6P} {641, A6}

\bibitem[\protect\citeauthoryear{{Pyne} \& {Joachimi}}{{Pyne} \&
  {Joachimi}}{2021}]{Pyne}
{Pyne} S.,  {Joachimi} B.,  2021, \mn@doi [\mnras] {10.1093/mnras/stab413},
  \href {https://ui.adsabs.harvard.edu/abs/2021MNRAS.503.2300P} {503, 2300}

\bibitem[\protect\citeauthoryear{{Reid}, {Seo}, {Leauthaud}, {Tinker}  \&
  {White}}{{Reid} et~al.}{2014}]{Reid14}
{Reid} B.~A.,  {Seo} H.-J.,  {Leauthaud} A.,  {Tinker} J.~L.,   {White} M.,
  2014, \mn@doi [\mnras] {10.1093/mnras/stu1391}, \href
  {https://ui.adsabs.harvard.edu/abs/2014MNRAS.444..476R} {444, 476}

\bibitem[\protect\citeauthoryear{{Richard} et~al.,}{{Richard}
  et~al.}{2019}]{4MOST}
{Richard} J.,  et~al., 2019, \mn@doi [The Messenger] {10.18727/0722-6691/5127},
  \href {https://ui.adsabs.harvard.edu/abs/2019Msngr.175...50R} {175, 50}

\bibitem[\protect\citeauthoryear{{Singh}, {Mandelbaum}  \& {More}}{{Singh}
  et~al.}{2015}]{Singh}
{Singh} S.,  {Mandelbaum} R.,   {More} S.,  2015, \mn@doi [\mnras]
  {10.1093/mnras/stv778}, \href
  {https://ui.adsabs.harvard.edu/abs/2015MNRAS.450.2195S} {450, 2195}

\bibitem[\protect\citeauthoryear{{Singh}, {Yu}  \& {Seljak}}{{Singh}
  et~al.}{2021}]{Singh21}
{Singh} S.,  {Yu} B.,   {Seljak} U.,  2021, \mn@doi [\mnras]
  {10.1093/mnras/staa3263}, \href
  {https://ui.adsabs.harvard.edu/abs/2021MNRAS.501.4167S} {501, 4167}

\bibitem[\protect\citeauthoryear{{Taruya} \& {Okumura}}{{Taruya} \&
  {Okumura}}{2020}]{Taruya20}
{Taruya} A.,  {Okumura} T.,  2020, \mn@doi [\apjl] {10.3847/2041-8213/ab7934},
  \href {https://ui.adsabs.harvard.edu/abs/2020ApJ...891L..42T} {891, L42}

\bibitem[\protect\citeauthoryear{{Taruya}, {Saito}  \& {Nishimichi}}{{Taruya}
  et~al.}{2011}]{Taruya11}
{Taruya} A.,  {Saito} S.,   {Nishimichi} T.,  2011, \mn@doi [\prd]
  {10.1103/PhysRevD.83.103527}, \href
  {https://ui.adsabs.harvard.edu/abs/2011PhRvD..83j3527T} {83, 103527}

\bibitem[\protect\citeauthoryear{{Taylor}, {Kitching}, {Bacon}  \&
  {Heavens}}{{Taylor} et~al.}{2007}]{Taylor}
{Taylor} A.~N.,  {Kitching} T.~D.,  {Bacon} D.~J.,   {Heavens} A.~F.,  2007,
  \mn@doi [\mnras] {10.1111/j.1365-2966.2006.11257.x}, \href
  {https://ui.adsabs.harvard.edu/abs/2007MNRAS.374.1377T} {374, 1377}

\bibitem[\protect\citeauthoryear{{Verde}, {Heavens}, {Percival}  \&
  {Matarrese}}{{Verde} et~al.}{2002}]{Verde02}
{Verde} L.,  {Heavens} A.~F.,  {Percival} W.~J.,   {Matarrese} S.,  2002, arXiv
  e-prints, \href {https://ui.adsabs.harvard.edu/abs/2002astro.ph.12311V} {pp
  astro--ph/0212311}

\bibitem[\protect\citeauthoryear{{Vlah}, {Chisari}  \& {Schmidt}}{{Vlah}
  et~al.}{2020}]{Vlah}
{Vlah} Z.,  {Chisari} N.~E.,   {Schmidt} F.,  2020, \mn@doi [\jcap]
  {10.1088/1475-7516/2020/01/025}, \href
  {https://ui.adsabs.harvard.edu/abs/2020JCAP...01..025V} {2020, 025}

\bibitem[\protect\citeauthoryear{{Yuan}, {Garrison}, {Eisenstein}  \&
  {Wechsler}}{{Yuan} et~al.}{2022}]{Yuan22}
{Yuan} S.,  {Garrison} L.~H.,  {Eisenstein} D.~J.,   {Wechsler} R.~H.,  2022,
  \mn@doi [\mnras] {10.1093/mnras/stac1830}, \href
  {https://ui.adsabs.harvard.edu/abs/2022MNRAS.515..871Y} {515, 871}

\bibitem[\protect\citeauthoryear{{Zaldarriaga} \& {Seljak}}{{Zaldarriaga} \&
  {Seljak}}{1997}]{Zaldarriaga}
{Zaldarriaga} M.,  {Seljak} U.,  1997, \mn@doi [\prd]
  {10.1103/PhysRevD.55.1830}, \href
  {https://ui.adsabs.harvard.edu/abs/1997PhRvD..55.1830Z} {55, 1830}

\bibitem[\protect\citeauthoryear{{Zhai} et~al.,}{{Zhai} et~al.}{2022}]{Zhai22}
{Zhai} Z.,  et~al., 2022, arXiv e-prints, \href
  {https://ui.adsabs.harvard.edu/abs/2022arXiv220308999Z} {p. arXiv:2203.08999}

\bibitem[\protect\citeauthoryear{{Zhang}, {D'Amico}, {Senatore}, {Zhao}  \&
  {Cai}}{{Zhang} et~al.}{2022}]{Zhang22}
{Zhang} P.,  {D'Amico} G.,  {Senatore} L.,  {Zhao} C.,   {Cai} Y.,  2022,
  \mn@doi [\jcap] {10.1088/1475-7516/2022/02/036}, \href
  {https://ui.adsabs.harvard.edu/abs/2022JCAP...02..036Z} {2022, 036}

\bibitem[\protect\citeauthoryear{{d'Amico}, {Gleyzes}, {Kokron}, {Markovic},
  {Senatore}, {Zhang}, {Beutler}  \& {Gil-Mar{\'\i}n}}{{d'Amico}
  et~al.}{2020}]{damico20}
{d'Amico} G.,  {Gleyzes} J.,  {Kokron} N.,  {Markovic} K.,  {Senatore} L.,
  {Zhang} P.,  {Beutler} F.,   {Gil-Mar{\'\i}n} H.,  2020, \mn@doi [\jcap]
  {10.1088/1475-7516/2020/05/005}, \href
  {https://ui.adsabs.harvard.edu/abs/2020JCAP...05..005D} {2020, 005}

\bibitem[\protect\citeauthoryear{{de Mattia} et~al.,}{{de Mattia}
  et~al.}{2021}]{deMattia21}
{de Mattia} A.,  et~al., 2021, \mn@doi [\mnras] {10.1093/mnras/staa3891}, \href
  {https://ui.adsabs.harvard.edu/abs/2021MNRAS.501.5616D} {501, 5616}

\bibitem[\protect\citeauthoryear{{van Gemeren} \& {Chisari}}{{van Gemeren} \&
  {Chisari}}{2021}]{vanGemeren}
{van Gemeren} I.~R.,  {Chisari} N.~E.,  2021, \mn@doi [\prd]
  {10.1103/PhysRevD.104.069902}, \href
  {https://ui.adsabs.harvard.edu/abs/2021PhRvD.104f9902V} {104, 069902}

\makeatother
\end{thebibliography}


\onecolumn
\appendix

\section{Clustering and alignment multipoles}
\label{app:multi}

Using Eq. (\ref{eq:multi}) we can compute the multipole power spectra for our observables of interest. The two lowest order multipoles for galaxy clustering (Eq. \ref{eq:pkggA}) are
\begin{eqnarray}
P_{gg}^{(0)}(k)&=&\left[b_g^2+\frac{2}{3}b_gf+\frac{1}{5}f^2+\frac{4}{45}(R\tilde{C}_1)^2+\frac{8}{45}fR\tilde{C}_1\right]P_\delta(k),\\
P_{gg}^{(2)}(k)&=&\left[\frac{4}{3}b_gf+\frac{4}{7}f^2+\frac{8}{63}(R\tilde{C}_1)^2+\frac{4}{3}b_gR\tilde{C_1}+\frac{44}{63}fR\tilde{C_1}\right]P_\delta(k),
\end{eqnarray}
where for simplicity we defined $R=-3.48(\eta\chi)_{\rm eff}$.
And the lowest order multipoles for the cross-correlation of galaxy positions and intrinsic shapes (Eq. \ref{eq:PgE}), and the auto-correlation of galaxy shapes (Eq. \ref{eq:PEE}) are, respectively,
\begin{eqnarray}
P_{gE}^{(0)}(k)&=&-\tilde{C}_1\left[\frac{2}{3}b_g+\frac{2}{15}f\right]P_\delta(k)\\
P_{EE}^{(0)}(k)&=&\frac{8}{15}\tilde{C}_1^2P_\delta(k)
\end{eqnarray}
Because we stay at linear order, all these expressions are analytically differentiable with respect to the parameters $\theta_i=\{b_g,f,\tilde{C}_1\}$. Such derivatives are needed to compute the Fisher matrix in Eq. (\ref{eq:fisher}).


\bsp	
\label{lastpage}
\end{document}